\documentclass[10pt]{wlscirep}
\usepackage[utf8]{inputenc}
\usepackage[T1]{fontenc}
\usepackage{tikz}
\usepackage[most]{tcolorbox}
\usepackage[outline]{contour}
\usetikzlibrary{quantikz2}
\usepackage{amsmath, amssymb}
\usepackage{braket}
\usepackage{adjustbox}
\usepackage{xcolor}
\usepackage{comment}
\usepackage{float}
\usepackage{subcaption}
\usepackage{lineno}

\title{A 1-bit quantum filter for particle trajectory reconstruction}

\author[1,2,3,*]{Xenofon Chiotopoulos}
\author[2]{Davide Nicotra}
\author[2,4]{George Scriven}
\author[3]{Kurt Driessens}
\author[1,2]{Marcel Merk}
\author[4]{Jochen Sch\"{u}tz}
\author[2]{Jacco de Vries}
\author[3]{Mark H.M. Winands}
\affil[1]{Nikhef National Institute for Subatomic Physics, Science Park 105, 1098 XG Amsterdam, The Netherlands}
\affil[2]{Maastricht University, Faculty of Science and Engineering, Gravitational Waves and Fundamental Physics department, Duboisdomain 30, 6229 GT Maastricht, The Netherlands}
\affil[3]{Maastricht University, Faculty of Science and Engineering, Department of Advanced Computing Sciences,  Paul-Henri Spaaklaan 1, 6229 EN Maastricht, The Netherlands} 
\affil[4]{Hasselt University, Faculty of Sciences and Data Science Institute, Agoralaan gebouw D, 3590 Diepenbeek, Belgium}

\affil[*]{xenofon.chiotopoulos@maastrichtuniversity.nl}

\begin{abstract}
The transition to the High-Luminosity Large Hadron Collider (HL-LHC) presents a computational challenge where particle reconstruction complexity may outpace classical computing resources. While quantum computing offers potential speedups, standard algorithms like Harrow-Hassidim-Lloyd (HHL) require prohibitive circuit depths for near-term hardware. Here, we introduce a 1-Bit Quantum Filter, a domain-specific adaptation of HHL that reformulates tracking from matrix inversion to binary ground-state filtering. By replacing high-precision phase estimation with a single-ancilla spectral threshold and exploiting the Hamiltonian's sparsity, we achieve an asymptotic gate complexity of $\mathcal{O}(\sqrt{N} \log N)$, given Hamiltonian dimension $N$. We validate this approach on LHCb Monte Carlo events, demonstrating segment finding efficiency highly competitive with the classical state-of-the-art methods. Furthermore, we benchmark performance using the Quantinuum System Model H2 trapped-ion processor and IBM Heron R3 superconducting processor. This work establishes a quantum track reconstruction method capable of solving realistic event topologies on noise-free simulators and smaller tracking scenarios within the current constraints of the Noisy Intermediate Scale Quantum (NISQ) era. Remaining challenges toward a full end-to-end tracking solution include an efficient readout and Hamiltonian construction.
\end{abstract}

\begin{document}

\flushbottom
\maketitle
\thispagestyle{empty}

\section*{Introduction}

Reconstructing particle trajectories from detector hits is a central aspect of experimental high energy physics. The work in this paper anticipates on a hybrid classical-quantum implementation of particle track-finding for the LHCb experiment in LHC Run-5~\cite{2019_HEP_COMUPTE_ROADMAP}, currently planned to start in 2036. In the High-Luminosity phase of the Large Hadron Collider (HL-LHC) at CERN, high-energy collisions will result in events where thousands of particles are simultaneously produced. These particles traverse multiple sensitive detection layers where they deposit small amounts of energy, generating discrete signals known as detector hits, from which the particle trajectories are reconstructed for physics analysis. The reconstruction process involves a significant combinatorial pattern recognition task, which must be solved in real-time at the high beam crossing rate of 40 MHz and requires expensive computational resources.

The transition to the HL-LHC presents a computational bottleneck that challenges the scalability of traditional track reconstruction algorithms. As the number of simultaneous collisions increases by an order of magnitude, the combinatorial complexity of track reconstruction grows factorially~\cite{search_by_triplet}, potentially outpacing the growth of the classical computing budgets. To address this, the high-energy physics community is increasing its exploration of disruptive computing paradigms, including machine learning and quantum computing. For instance, Graph Neural Networks (GNNs) have shown potential but require extensive training and their inference latency goes beyond the strict processing constraints of the real-time LHCb trigger~\cite{Ju_2021_GNN, Correia_2024_GNN}. Quantum computers may offer advantages including potential better energy consumption efficiency \cite{Jaschke_2023_green, PRXQuantum.3.020101_green} and the ability to leverage high-dimensional Hilbert spaces to address combinatorial problems. Various approaches for particle tracking have been explored, ranging from Quantum Annealing~\cite{Zlokapa_2021, quantum_annealing} to quantum GNNs~\cite{QGNN}. In previous work we explored using the Harrow-Hassidim-Lloyd (HHL~\cite{Harrow_2009}) algorithm as a global pattern recognition algorithm, TrackHHL~\cite{Nicotra_2023,1-Bit_HHL}, achieving good performance but requiring circuit depths that far exceed the coherence times available on current Noisy Intermediate-Scale Quantum (NISQ) devices. 

In this paper we present an extension to TrackHHL\cite{1-Bit_HHL}, which introduces a specialized 1-Bit Quantum Filter to mitigate the limitations of HHL. Inspired by the Denby-Peterson model~\cite{DP1, DP2}, track reconstruction is formulated as the ground-state search of an Ising-like Hamiltonian, which we approach as a spectral filtering task rather than a resource-intensive, high-precision matrix inversion. By restricting Quantum Phase Estimation (QPE) to a single-bit clock register and exploiting phase aliasing through a problem-informed evolution parameter, the algorithm deterministically filters out combinatorial noise while constructing the tracking solution. We leverage the Hamiltonian's sparsity and structure by implementing a custom time evolution using the proposed Direct Structural Synthesis (DSS) method, which synthesizes exact unitary operators to bypass Trotterization and reduce circuit depth. This approach achieves an analytical two-qubit gate complexity of $\mathcal{O}(\sqrt{N} \log N)$, providing a polynomial speedup over classical matrix inversion and an overall sampling complexity of $\mathcal{O}(N \log N)$, where $N$ is the number of candidate track segments representing the dimension of the linear system. We validate our 1-Bit Quantum Filter using noiseless statevector simulations on both fast MC-toy events and on realistic LHCb Monte Carlo (MC) simulated events containing up to 1,000 hits, demonstrating a segment-finding efficiency of $94.2\%$, which is comparable with the classical state-of-the-art "Search by Triplet" algorithm~\cite{search_by_triplet}. Finally, we benchmark the algorithm's noise resilience through physical executions on IBM's Heron R3 superconducting processors~\cite{AbuGhanem_2025_ibm_heron} and Quantinuum's System Model H2 trapped-ion system~\cite{Moses_2023_race_track}. We demonstrate that for our approach the all-to-all connectivity and lower error rates of the trapped-ion architecture provide improved solution fidelity for small systems with no error correction, while both systems currently suffer from intractable gate error accumulation beyond a problem size of 4 particle tracks and 3 layers. These findings establish a viable pathway to further study quantum particle track reconstruction at the scale of future HL-LHC events.

\section*{Methods}

\subsection*{Hamiltonian Formulation for Trackfinding}
\label{sec:Hamiltonian}

\begin{figure}[H]
	\centering
    \includegraphics[width=17.0cm]{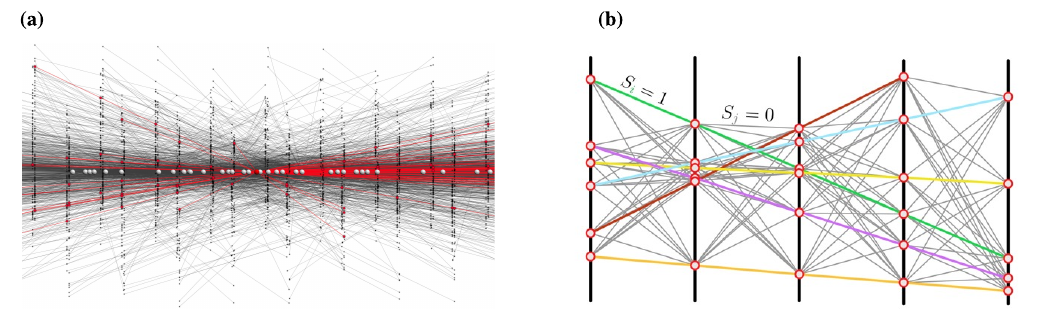}
    \caption{\textbf{ Event simulation mapping to the tracking graph.} \textbf{(a)} A simulated event in the LHCb detector. The larger grey dots represent the primary vertices of multiple simultaneous proton-proton collisions, while the small black dots indicate the detector hits generated by the produced particles. The faint grey lines depict the reconstructed tracks of the produced particles. The red dots and lines indicate the hits and particles produced in a single given collision, represented by the red circle. \textbf{(b)} Illustration of the definition of the Hamiltonian graph construction~\cite{Nicotra_2023}. Coloured segments represent active variables $S_i=1$ forming valid tracks, while grey segments correspond to inactive variables $S_j=0$.}
     \label{fig:Event}
\end{figure}

The LHCb \textbf{Ve}rtex \textbf{Lo}cator (VELO) \cite{LHCBUpgrade} serves as the innermost tracking detector in the spectrometer, positioned 5 mm from the proton-proton interaction point. It comprises 52 silicon pixel detector modules, divided equally between two retractable halves surrounding the beam line. At the High-Luminosity LHC each beam crossing causes $\sim 20$ simultaneous collisions, which lead to $\sim 1000$ produced particles along straight line trajectories due to the absence of a magnetic field. When charged particles traverse the detector modules they deposit energy, allowing the VELO to record precise two-dimensional hit coordinates $(x, y)$ and derive a $z$ coordinate from the detector module position. The subsequent reconstruction process aims to group these hits into distinct particle trajectories (tracks). One such simulated beam crossing event in the LHCb VELO is shown in the left panel of Fig.~\ref{fig:Event}. To convert the task of track reconstruction into a format suitable for quantum processing, we formulate the problem within a Hamiltonian framework, adapting the approach initially proposed by Denby and Peterson\cite{DP1,DP2} and subsequently refined for the LHCb VELO geometry \cite{Nicotra_2023, 1-Bit_HHL}, briefly summarized below. 

The system state is parameterized by binary variables $\textbf{S} = \{S_i, 1 \leq i \leq N, S_i \in \{0,1\}\}$, representing $N$ potential track segments (doublets connecting hits on adjacent detector layers), where $S_i = 1$ indicates an active segment contributing to a track, and $S_i = 0$ indicates inactivity. This graph representation is illustrated in the right panel of Fig.~\ref{fig:Event}. The task of identifying valid tracks is then mapped to finding the ground state of the Hamiltonian $\mathcal{H}(\textbf{S})$ with angular, spectral and gap terms, given by:
\begin{equation}
    \mathcal{H}(\mathbf{S}) = \mathcal{H}_{\text{ang}}(\mathbf{S},\epsilon) + \alpha \mathcal{H}_{\text{spec}}(\mathbf{S}) + \beta \mathcal{H}_{\text{gap}}(\mathbf{S}) \; .
    \label{eq:DP-hamiltonian} 
\end{equation}
Here, $\mathcal{H}_{\text{ang}}$ rewards straight connections between segments ${S}_{i}$ and ${S}_{j}$ where the angle they construct $\theta_{i,j}$ is smaller than some angular tolerance $\epsilon$:
\begin{equation}
    \mathcal{H}_{\text{ang}} (\mathbf{S}, \epsilon) = -\frac{1}{2} \sum_{ i,j } f(\theta_{i,j}, \epsilon) S_{i} S_{j}, \quad
    f(\theta, \epsilon) = \begin{cases}  1 & \text{if } \cos \theta \geq 1 - \epsilon, \\ 0 & \text{otherwise} \; . \end{cases} 
    \label{eq:angular_term} 
\end{equation}

The terms $\mathcal{H}_{\text{spec}}(\mathbf{S}) = \sum_{i}{S}^2_{i}$ and $\mathcal{H}_{\text{gap}}(\mathbf{S}) = \sum_{i}(1-{S}_{i})^2$ serve as regularization constraints, weighted by parameters $\alpha = 2.0$ and $\beta = 1.0$ following the previous work \cite{Nicotra_2023}. To leverage quantum linear solvers, this discrete optimization problem is transformed into a linear system, following the mathematical framework established for quantum Hopfield neural networks~\cite{PhysRevA.98.042308_hopfield_quantum}. By relaxing the discrete segment variables ${S}_i$ to continuous variables forming a real-valued vector $\mathbf{x} \in \mathbb{R}^N$, the minimum of $\mathcal{H}(\mathbf{x})$ can be found by setting its gradient to zero:
\begin{equation}
    \nabla_{\mathbf{x}} \mathcal{H} = \mathbf{A} \mathbf{x} - \mathbf{b} = 0 \quad \implies \quad \mathbf{A} \mathbf{x} = \mathbf{b}  \quad  \implies  \quad  \mathbf{x} = \mathbf{A}^{-1}\mathbf{b} \; .
    \label{eq:relaxation_derivative}
\end{equation}
The resulting matrix $\mathbf{A}$ encapsulates the quadratic couplings defined in $\mathcal{H}$. The vector $\mathbf{b}$ originates from the linear terms in $\mathbf{S}$ (arising from $\mathcal{H}_{\text{gap}}$) and takes the form $\mathbf{b}=(1,1,\dots,1)^T$ due to the choice of $\beta = 1$. The parameters $\alpha$ and $\beta$ are chosen such that the resulting matrix is positive semi-definite. The solution $\mathbf{x}$ to this linear system represents the relaxed segment amplitudes, which must then be discretized (applying a threshold $T$ via sampling high-amplitude segments) to identify the final set of track segments. This formulation maps the combinatorial track-finding problem onto a linear algebraic structure compatible with algorithms like HHL.

To quantify the computational scale of this approach, we define the system parameters in terms of the detector geometry and event size. The dimension of the linear system, $N$, corresponds to the total number of candidate segments in the set $\mathbf{S}$, where $N\times N$ defines the size of the $\mathbf{A}$ matrix~\cite{Nicotra_2023}. For a detector with $l$ layers and $m$ reconstructible particle tracks, the system size scales as $N=m^2(l-1)$. The total number of physical interaction terms, denoted by $k$, represents the number of angular constraints active in the system, and defines the number of non-zero off-diagonal elements in the Hamiltonian. For full tracks spanning the detector, this scales linearly with the number of particles as $k = m(l-2)$. These parameters will define the gate complexity and sampling cost of our proposed quantum algorithm. 
 
\subsection*{Simulation Tool}
\label{sec:simulation_tool}

To evaluate the performance of the reconstruction algorithm and in particular to explore the spectral and structural properties of the Hamiltonian formulation in Eq.~(\ref{eq:DP-hamiltonian}), we developed a dedicated simulation tool. This tool captures the essential geometric and detection features of the LHCb VELO, such as particle multiple scattering, detector resolution as well as hit (in-)efficiency, while remaining computationally tractable for hybrid quantum-classical studies.

\begin{figure}[H]
    \centering
    \includegraphics[width=17.0cm]{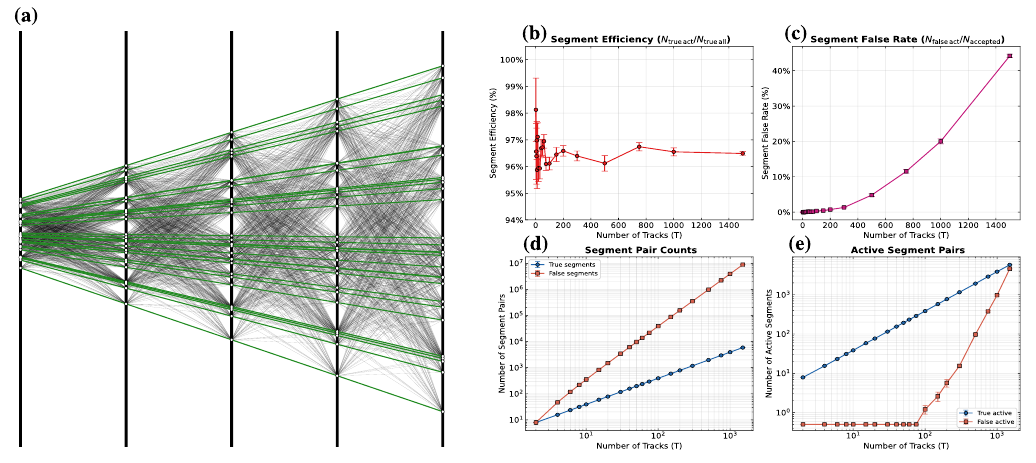}
    {\phantomsubcaption\label{fig:ToyEvent}
    \phantomsubcaption\label{fig:ToyPerformanceB}\
    \phantomsubcaption\label{fig:ToyPerformanceC}
    \phantomsubcaption\label{fig:ToyPerformanceD}
    \phantomsubcaption\label{fig:ToyPerformanceE}}
    \caption{\textbf{Fig. 2: Tracking performance metrics of the classical matrix solver.} \textbf{(a)} Visual event display of a simulated collision with a single primary vertex, 5 detection layers, and 64 particle tracks originating from a single primary vertex. The valid track segments are shown as green lines, and the inactive combinatorial background segments are shown as black lines. \textbf{(b,c,d,e)} Classical-solver segment performance versus track multiplicity $T$ at fixed angular tolerance $\varepsilon=2$~mrad with a 1\% random hit inefficiency. Markers have vertical error bars representing the standard error of the mean ( $\mathrm{SEM}=\sigma/\sqrt{n}$) across independent toy events at each $T$ ($n=60$ events for $T\le100$, $n=30$ for $150\le T\le300$, $n=15$ for $T\ge500$). \textbf{(b)} Segment efficiency, the fraction of true segments activated by the solver; red circles show the average as function of the number of tracks in the event. \textbf{(c)} Segment false rate, the fraction of incorrectly activated segments, shown as magenta squares. \textbf{(d)} The number of candidate segments prior to solving: blue circles are true segments, red-orange squares are false (ghost) segments. \textbf{(e)} The segments activated by the solver,  for true active (blue-circle) and false active segments (red-orange-square) respectively.}
    \label{fig:SimulationOverview}
\end{figure}

The tool implements a given number of sequential measurement planes in $z$ that register hits of traversing particle tracks produced in events. The events consist of a configurable number of linear particle tracks originating from one or more collision vertices.
Given the initial vertex position $(x=0,\; y=0,\; z=z_0)$, each track is parametrized by a state vector $(x,\; y,\; t_x,\; t_y,\; p/q)_z$ at position $z$, where $t_x = dx/dz$ and $t_y = dy/dz$ denote the slopes of the trajectory in the transverse plane, and $p/q$ is the charge--momentum ratio, included for completeness as in the absence of a magnetic field the particles propagate linearly between the detector layers. At each detector layer $z=z_i$ the track deposits a {\em hit}, which is registered with an experimental realistic Gaussian measurement resolution. Subsequently, the track undergoes multiple scattering slightly changing its slopes $t_x$ and $t_y$ according to Gaussian multiple scattering $\sigma_{\text scatt}$ of the particle in the material. The average scattering angle implemented is given by the material thickness of the detection planes and is inversely proportional to the momentum of the traversing particle, and therefore allows to study the performance as function of particle track momentum. 
Fig.~\ref{fig:ToyEvent} illustrates an event as generated from a single collision point.

A classical algorithm using classical matrix inversion techniques is used to test the potential of the Hamiltonian approach on the toy. Its performance is evaluated as a function of the number of collisions with the segment-finding {\em efficiency}: the fraction of correctly found track segments, and the {\em fake rate}: the fraction of fake segments in the total sample. Figs.~\ref{fig:ToyPerformanceB},~\ref{fig:ToyPerformanceC},~\ref{fig:ToyPerformanceD},~\ref{fig:ToyPerformanceE} summarize the performance of the method in terms of segment finding efficiency $\epsilon_\text{segment}$ and fake rate $f_\text{segment}$ with:

\begin{equation}
    \epsilon_\text{segment}=\frac{N_\text{true accepted}}{N_\text{true generated}} \hspace*{1.0cm}
    \text{and} \hspace*{1.0cm}
    f_\text{segment}=\frac{N_\text{false accepted}}{N_\text{all accepted}}
    \label{eq:segment_eff}
\end{equation}

With the current Hamiltonian formulation, the efficiency remains close to 100\%, independent of problem size.
For high track densities the algorithm shows an increasing fake rate. Future work is envisaged to include additional bifurcation and occupancy terms from the Denby-Peterson model, as well as an improved angular term modeling competition between segment candidates.

\subsection*{Quantum Algorithm}

As described in the Methods section, this work builds upon the quantum tracking framework established in Ref.~\cite{Nicotra_2023,1-Bit_HHL}, which uses the HHL~\cite{Harrow_2009} algorithm to solve linear systems of equations described by the Hamiltonian track finding formulation in Eq.~\eqref{eq:DP-hamiltonian}.  HHL's core advantage comes from the Quantum Phase Estimation (QPE~\cite{Cleve_1998_qpe, kitaev1995quantummeasurementsabelianstabilizer_qpe}) subroutine to resolve these eigenvalues into a multi-qubit clock register. QPE consists of a Hamiltonian evolution followed by an inverse Quantum Fourier Transform (QFT~\cite{coppersmith2002approximatefouriertransformuseful_qft}). The QPE's required precision determines the number of applications of the controlled unitary that implements the time evolution. A controlled rotation is then applied to an ancilla qubit to encode the inverse eigenvalues ($1/ \lambda$) onto the state amplitudes, followed by the uncomputation of the clock register. HHL is theoretically powerful and promising a logarithmic speedup, though its reliance on a high-precision estimation of the spectrum coupled with an expensive tomography creates significant bottlenecks~\cite{Aaronson2015_HHL_challenges, dervovic2018quantumlinearsystemsalgorithms_HHL_challenges}. The algorithm requires prohibitive circuit depths due to iterative application of the Hamiltonian evolution $e^{-iAt}$ and QPE requires many clock register qubits, rendering it unfeasible for particle tracking on near-term hardware. 

To address these limitations we present a 1-Bit Quantum Filter, a formalization and extension of our previous work~\cite{1-Bit_HHL}. We observe that for track reconstruction, exact matrix inversion is not necessary. The tracking problem is inherently binary, the task is to distinguish valid 'signal' segments from 'noise' segments. Consequently, we replace the resource-intensive high-precision QPE and controlled rotation using a 1-bit QPE with a deterministic eigenvalue filter. This reformulates the task from exact inversion to ground-state spectral filtering, drastically reducing the gate complexity and sampling requirements~\cite{1-Bit_HHL}. Additionally, we replace the general purpose approximate Suzuki-Trotter evolution~\cite{Suzuki1976} with a Direct Structural Synthesis (DSS) method, which synthesizes controlled unitary operators without the need for an expensive classical decomposition.  
 
\subsubsection*{A 1-Bit Quantum Filter}

\begin{figure}[htbp]
    \includegraphics[width=17.0cm]{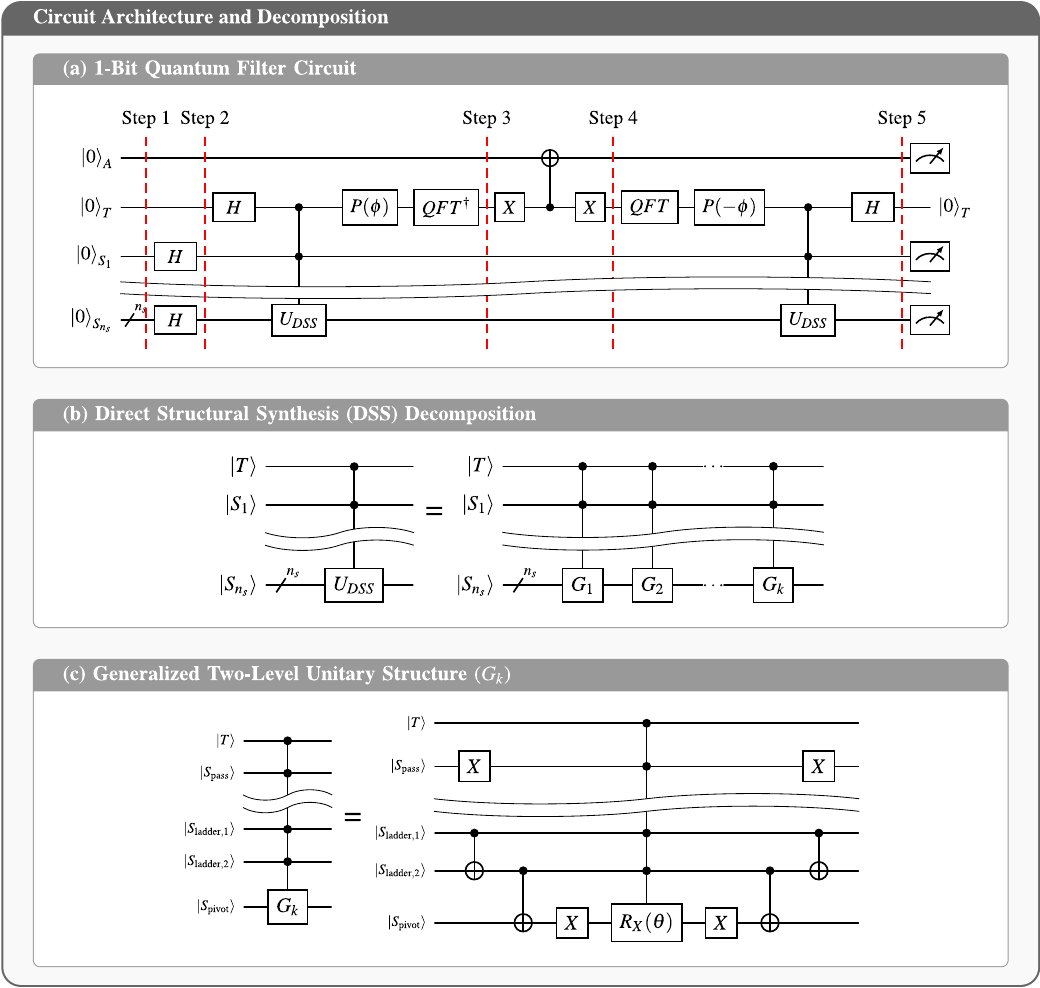}
    {
    \phantomsubcaption\label{fig:combined_circuit:a}
    \phantomsubcaption\label{fig:combined_circuit:b}
    \phantomsubcaption\label{fig:combined_circuit:c}
    }
    \caption{\textbf{Fig. 3: Circuit Architecture.} \textbf{(a)} Circuit diagram of the proposed 1-Bit Quantum Filter, with registers that include the single ancilla flag $\ket{0}_A$, the single time register $\ket{0}_T$ for the 1-bit QPE, and the system register $\ket{0}_{S_1} \dots\ket{0}_{S_{n_s}}$, where $n_s=log_2N$ represents the total number of system qubits. The algorithm is composed of state preparation, 1-bit phase estimation with flagging logic, and uncomputation. The additional phase gate $P(\phi)$ on the time register implements the diagonal term of the Hamiltonian. \textbf{(b)} The Direct Structural Synthesis (DSS) decomposition, explicitly showing that the controlled time evolution $U_{\text{DSS}}$ is implemented as a sequential application of controlled interaction gates $\prod_{k} G_k = \prod_{k} e^{i B_k t}$ corresponding to the non-zero off-diagonal terms of the Hamiltonian matrix. \textbf{(c)} The generalized structure of a single interaction gate $G_k = e^{i B_k t}$. For states with Hamming distance $d_H > 1$, a CNOT ladder ($S_{\text{lad}}$) disentangles the states onto a pivot ($S_{\text{piv}}$), while passive qubits ($S_{\text{pass}}$) act as controls to strictly enforce subspace selection.}
    \label{fig:combined_circuit}
\end{figure}

This approach replaces resource-intensive eigenvalue inversion with a binary filtration mechanism that takes advantage of the specific spectral properties of the track-finding Hamiltonian.The algorithm begins with determining the evolution time $t$, which is used in our 1-bit phase estimation subroutine. In general, the spectrum bounds $(\lambda_{\text{min}}, \lambda_{\text{max}})$ can be estimated via the Gershgorin Circle Theorem~\cite{Horn_Johnson_1985}. However, for the particle tracking case the bounds are known as they converge to $(\lambda_{\text{min}} \approx 1, \lambda_{\text{max}} \approx 5)$ \cite{Nicotra_2023}. Since the physical length of a track is bounded by the number of detector layers ($l$), the eigenvalues of the track sub-matrices are bounded by the track length rather than the total number of candidate segments $N$. Due to this convergence, we can target the central eigenvalue $\lambda_c$. All the noise states exist in the eigenspace associated with the central eigenvalue, the numerical value of this is determined by $\alpha + \beta$ from Eq.~\eqref{eq:DP-hamiltonian}:
\begin{equation}
    \lambda_c = \alpha + \beta \quad \text{and} \quad t = \frac{\pi}{\lambda_c} \; .
    \label{eq:evo_time}
\end{equation}
This choice of $t$ is a deliberate deviation from standard QPE, where the evolution time is typically scaled as a function of $\lambda_{max}$, to prevent phase aliasing \cite{Harrow_2009, Nielsen_Chuang_2010}. We instead reduce the precision of the QPE to a single time register qubit and take advantage of the aliasing due to our choice of $t$ in Eq. \eqref{eq:evo_time}, which allows us to project the noise to a specific phase. We prepare three registers: the system register $\ket{0}_S$ for encoding the $\mathbf{b}$ vector, the time register used for the QPE $\ket{0}_T$ and the ancilla register used to determine if the algorithm has run successfully $\ket{0}_\mathbf{A}$. The system register (comprising $n_s=log_2N$ qubits) is initialized in the uniform superposition state $\ket{b} = H^{\otimes n_s} \ket{0} = \ket{+}^{\otimes n_s}$, as $\mathbf{b}$ is defined as a uniform vector. To analyze the spectral filtering mechanism, we rewrite the state in the eigenbasis of the Hamiltonian matrix $\mathbf{A}$:
\begin{equation}
    \quad \ket{b} = \sum_j \gamma_j \ket{u_j} \; ,
    \label{eq:eigenbasis_decomopition}
\end{equation}
where $\ket{u_j}$ are the orthonormal eigenvectors with eigenvalues $\lambda_j$, and $\gamma_j = \braket{u_j|b}$ represents the overlap amplitude. The proposed 1-Bit Quantum Filter proceeds in five steps, tracking the evolution of the state $\ket{\Psi}$ across the ancilla $\ket{A}$, time $\ket{T}$, and system $\ket{S}$ registers. The following steps match exactly the circuit shown in Fig.~\ref{fig:combined_circuit:a}.

\begin{description}
    \item[Step 1: State Preparation] 
    We initialize the system register $\ket{0}_S$ into a uniform superposition $\ket{b} = \ket{+}^{\otimes n_s}$, representing the vector $\mathbf{b}=(1,1,\dots,1)^T$. To track the phase evolution in subsequent steps, we express the state $\ket{+}^{\otimes n_s}$ in the eigenbasis of the Hamiltonian matrix $\mathbf{A}$ as in Eq.~\ref{eq:eigenbasis_decomopition}. The initial state is:
    \begin{equation}
        \ket{\Psi_1} = \ket{0}_A \otimes \frac{1}{\sqrt{2}}(\ket{0} + \ket{1})_T \otimes \sum_j \gamma_j \ket{u_j}_S \; .
        \label{eq:step1}
    \end{equation}
    The time register is also prepared in superposition via a Hadamard gate.
    
    \item[Step 2: 1-Bit Phase Estimation] 
    We apply the controlled time evolution $U = e^{iAt}$ conditioned on the $\ket{1}_T$ state. This unitary is implemented by combining the Direct Structural Synthesis (DSS) method applied to non-diagonal terms, with a Phase Gate $P(\phi)$ (where $\phi = -(\alpha + \beta)t$) that accounts for the diagonal contributions of the Hamiltonian. Together, these components realize the time evolution of $A$, resulting in the phase $e^{i \lambda_j t}$ being kicked back onto the $\ket{1}_T$ state for each eigen-component:
    \begin{equation}
        \ket{\Psi_{2_a}}_j = \ket{0}_A \otimes \frac{1}{\sqrt{2}} \sum_j \gamma_j \left( \ket{0}_T + e^{i \lambda_j t} \ket{1}_T \right) \otimes \ket{u_j}_S \; . 
        \label{eq:step2a}
    \end{equation}
    This creates a relative phase between the $\ket{0}_T$ and $\ket{1}_T$ states. Next, the inverse QFT is applied to the time register $|0\rangle_T$. In the 1-Bit case the inverse QFT is implemented using a single Hadamard on the time register to interfere the phases. Using the identity $H(|0\rangle + e^{i\phi}|1\rangle) \varpropto \cos(\phi/2)|0\rangle - i\sin(\phi/2)|1\rangle$\cite{Nielsen_Chuang_2010}, it follows that:
    \begin{equation}
        \ket{\Psi_{2_b}}_j = \ket{0}_A \otimes \sum_j \gamma_j \left[ \cos\left(\frac{\lambda_j t}{2}\right) \ket{0}_T - i\sin\left(\frac{\lambda_j t}{2}\right)\ket{1}_T \right] \otimes \ket{u_j}_S \; .
        \label{eq:step2b}
    \end{equation}
    The phase information $\lambda_j t$ has been converted into amplitude information in the $|0\rangle_T$ and $|1\rangle_T$ basis states.

    \item[Step 3: Conditional Logic] 
    We apply a Zero-Controlled NOT gate on the ancilla, flipping $\ket{0}_A \to \ket{1}_A$ only when the time register is $\ket{0}_T$. The relative phase relationship is preserved, but now entangled with the ancilla register:
    \begin{equation}
        \ket{\Psi_3} = \sum_j \gamma_j \left[ \underbrace{\cos\left(\frac{\lambda_j t }{2}\right) \ket{1}_A \ket{0}_T}_{\text{Signal}} - \underbrace{i\sin\left(\frac{\lambda_j t }{2}\right)\ket{0}_A \ket{1}_T}_{\text{Noise}} \right] \otimes \ket{u_j}_S \; .
        \label{eq:step3}
    \end{equation}
    \item[Step 4: Uncomputation] 
    To disentangle the time register, we apply the inverse of the phase estimation circuit (Inverse QPE). This sequence consists of applying a Hadamard gate, the inverse controlled time evolution $C(e^{-i\mathbf{A}t})$, and a final Hadamard gate to the time register. Applying this sequence to $\ket{\Psi_3}$ yields the exact full intermediate state:

    \begin{equation}
        \ket{\Psi_{4a}} = \sum_j \gamma_j \left[ \left( \cos^2\frac{\lambda_j t}{2} \ket{1}_A + \sin^2\frac{\lambda_j t}{2} \ket{0}_A \right) \ket{0}_T + \frac{i}{2}\sin(\lambda_j t) ( \ket{1}_A - \ket{0}_A ) \ket{1}_T \right] \otimes \ket{u_j}_S \; .
        \label{eq:step4_intermediate}
    \end{equation}

    This explicit expansion demonstrates how the time register is restored. For the combinatorial noise subspace ($j \in Noise$), our specific choice of evolution time $t = \pi/\lambda_c$ ensures the time register perfectly uncomputes to $\ket{0}_T$ and is entirely isolated into the rejected $\ket{0}_A$ state. For the signal subspace ($\lambda_j \neq \lambda_c$), the uncomputation splits across the time register with minor phase leakage. The idealized target state projected onto the successfully uncomputed $\ket{0}_T$ subspace is grouped as:

    \begin{equation}
        \ket{\Psi_4} = \left[ \left( \ket{1}_A \otimes \sum_{j \in Signal} \gamma_j \ket{u_j}_S \right) + \left( \ket{0}_A \otimes \sum_{j \in Noise} \gamma_j \ket{u_j}_S \right) \right] \otimes \ket{0}_T \; .
        \label{eq:step4}
    \end{equation}

    \item[Step 5: Measurement \& Filtration] 
    Finally, we measure the ancilla. Our specific choice of $t = \pi / \lambda_c$ ensures that for noise eigenvalues ($\lambda = \lambda_c$), the probability of triggering the signal flag vanishes:
    \begin{equation}
        P(1 | \lambda_c) = \cos^4\left(\frac{\lambda_c t}{2}\right) + \frac{1}{4}\sin^2(\lambda_c t) = \cos^2\left(\frac{\lambda_c t}{2}\right) = \cos^2\left(\frac{\pi}{2}\right) = 0 \; .
        \label{eq:P1_succ}
    \end{equation}
    As a result, post-selecting on the measurement projecting onto the signal subspace $\ket{1}_A$ deterministically filters out the combinatorial noise. The collapsed state must then be renormalized by the probability of success $P_{succ}$, yielding the solution state:
    \begin{equation}
        \ket{\Psi_{5}} = \frac{1}{\sqrt{P_{\text{succ}}}} \sum_{j \in \text{Signal}} \gamma_j \cos^2\left(\frac{\lambda_j t}{2}\right) \ket{u_j}_S \; .
        \label{eq:step5}
    \end{equation}
\end{description}

This spectral separation is guaranteed by the Hamiltonian structure defined in Eq.~\eqref{eq:DP-hamiltonian}. Since noise segments correspond to isolated nodes in the graph, their eigenvalues are fixed exactly at $\lambda_c$, while valid track segments are spectrally shifted away. To exploit this, we replace the standard HHL rotation with a Zero-Controlled NOT gate (Step 3, Eq.~\ref{eq:step3}). This gate targets the ancilla conditioned on the $|0\rangle_T$ state, flipping it to $|1\rangle_A$ only for signal components. Consequently, noise components (which map to $|1\rangle_T$) fail to trigger the gate and remain in $|0\rangle_A$. Following the uncomputation step (Step 4, Eq.~\ref{eq:step4}), we measure the ancilla and post-select on the $|1\rangle_A$ outcome. This deterministically filters the noise contribution ($P(1|\lambda_c)=0$), leaving a final state composed strictly of valid track segments.

\subsubsection*{Time Evolution through Direct Structural Synthesis}
\label{sec:DSS}

A central component of our 1-bit Quantum Filter is the implementation of the controlled time evolution $C(U)=C(e^{-iAt})$. A general-purpose Suzuki-Trotter time evolution of $A$ is resource intensive \cite{1-Bit_HHL}, as it requires the $A$ matrix to be represented as a sum of Pauli strings. An alternative approach is the Direct Structural Synthesis (DSS) method that exploits the specific structure of our Hamiltonian, placing it within the class of exact gate synthesis algorithms \cite{Cowtan_2020_exact_synthesis}. The DSS method is based on the fact that we can express $A = cI - B$. Since the diagonal matrix term $cI$ commutes with the off-diagonal interaction matrix $B$, the unitary evolution operator $U(t) = e^{-iAt}$ allows for an exact factorization:
\begin{equation}
    U(t) = e^{-i(cI - B)t} = e^{-icIt} \cdot e^{iBt} \; .
    \label{eq:trotter_factorization}
\end{equation}
Implementing the controlled version of this evolution, $C(U)$, separates into a controlled global phase rotation ($C(e^{-icIt})$) and the controlled interaction terms ($C(e^{iBt})$).

The first term $C(e^{-icIt})$ can be efficiently implemented with a single $P(\phi)$ gate on the control qubit, with $\phi = -ct$. The phase rotation can be implemented with a single gate as $cI$ is a uniform identity matrix, its phase is kicked back deterministically from the target qubit to the control \cite{Nielsen_Chuang_2010}. The matrix $B=\sum_{k} B_{k}$ is a sum of its individual interaction components, where the index $k$ represents a unique interaction term $\{i,j\}$, and the total number of interaction terms that pass the $\mathcal{H}_{\text{ang}} (\mathbf{S}, \epsilon)$ angular function can be written as $k=m(l-2)$ for $m$ tracks. Each $B_{k}$ represents a $\sigma_x \text{-like}$ interaction between two basis states. The algorithm implements $C(e^{iB\textit{t}})$ by iterating through $k$ components and applying the gate for each one sequentially:
\begin{equation}
    G_{\text{DSS}} = \prod_{k} G_{k} \quad \text{where} \quad G_{k} = C(e^{iB_{k} {t}}) \; .
    \label{eq:dss_prod}
\end{equation}
A circuit based illustration of Eq. \eqref{eq:dss_prod} is visualized in Fig. \ref{fig:combined_circuit:b}. To strictly enforce the graph topology of $B$, each gate $G_{k}$ is implemented as an exact Two-Level Unitary, often referred to as a Givens rotation \cite{Nielsen_Chuang_2010}. For an interaction between basis states $\ket{i}$ and $\ket{j}$, the operator $G_k$ performs a rotation strictly within the subspace spanned by $\{ \ket{i}, \ket{j}\}$ and acts as the identity on all other states. This is synthesized using a ladder of CNOT gates to disentangle the basis states \cite{PhysRevA.103.042405_cnot_ladder}, followed by a multi-controlled rotation $C^{n-1}(R_X(\theta))$, and a subsequent uncomputation step.

The exactness of DSS time evolution hinges on whether the individual interaction terms $B_{k}$ commute. While this holds for terms acting on disjoint subspaces, the general case for a $d$-sparse Hamiltonian involves interaction terms sharing basis states. For these non-commuting terms, the factorization $e^{iB t} \neq \prod_{k} e^{iB_{k} t}$ is not an equality, but rather a first-order Suzuki-Trotter product formula\cite{Suzuki1976}: 
\begin{equation} 
    e^{iBt} = \prod_{k} e^{iB_{k} t} + \mathcal{O}(t^2) \; . 
    \label{eq:trotter_error} 
\end{equation} 
While this approximation introduces an error $\mathcal{O}(t^2)$ scaling quadratically with time, it is crucial to note that this error is subspace dependent. Specifically, the primary function of our filter remains robust because the noise eigenvectors $\ket{\psi_c}$ lie in the null space of the interaction matrix $B$. Since noise segments correspond to isolated vertices in the graph, they have no off-diagonal couplings, meaning $B\ket{\psi_c} = 0$ exactly. Consequently, the time evolution is exact for the noise subspace, ensuring that combinatorial background suppression is unaffected by time-evolution errors. The Trotter error $\mathcal{O}(t^2)$ acts non-trivially only on the signal eigenspace, introducing small phase perturbations that may slightly affect signal acceptance but not noise rejection.

While the noise subspace is immune to this error, the Trotter error acts non-trivially on the solution eigenspace ($\lambda_j$). The first-order simulation $U_{\text{sim}}(t)$ returns a perturbed phase $\phi_{\text{sim}} = \phi_{\text{exact}} + \delta\phi$, where $\phi_{\text{exact}} = \lambda_j t$ and $\delta\phi$ represents the $\mathcal{O}(t^2)$ phase error. Consequently, the measured probability for the solution state is shifted:
\begin{equation}
    P(0 | \lambda_j) =  \cos^2\left(\frac{(\lambda_j t + \delta\phi)}{2}\right) \quad \text{and} \quad P(1 | \lambda_j) =  \sin^2\left(\frac{(\lambda_j t + \delta\phi)}{2}\right) \; .
    \label{eq:supp_prob_error}
\end{equation}
The phase shift $\delta\phi$ perturbs the interference condition. While $\delta\phi$ does not cause spectral leakage for the critical $\lambda_c$ state, a large enough $\delta\phi$ can blur the signal phases $\lambda_j$. This can cause a portion of the solution states to be projected onto the reject bin ($|1\rangle_T$) by mistake, or conversely, shift the acceptance probability $P(0)$ closer to 1 or 0 depending on the sign of the error term. For the precision required in this application, the first-order approximation is sufficient; however, if higher precision QPE is critical, a higher-order product formula (e.g., 2nd-order Suzuki-Trotter with $\mathcal{O}(t^3)$ scaling) could be employed to suppress the error scaling.

\subsection*{Complexity Analysis}

To quantify the impact of the changes to HHL we study the computational complexity of the 1-Bit Quantum Filter and derive the asymptotic complexity for both two-qubit gate count and sampling cost. We demonstrate that the total gate and sampling complexity scales as $\mathcal{O}(\sqrt{N} \log N)$ and $\mathcal{O}(N \log N)$, respectively.

\subsubsection*{Gate Complexity}

We derive the asymptotic gate complexity for a single execution of the circuit. Based on the system parameters defined in Hamiltonian construction, the number of interaction terms scales as $k \approx \sqrt{N}$. Consequently, the total complexity scales as $\mathcal{O}(\sqrt{N} \log N)$, dominated by the sequential implementation of these off-diagonal terms in the time evolution. The total gate complexity $C_{gate}$ is the sum of the complexity in steps 1-4 in Fig. \ref{fig:combined_circuit:a}:

\begin{enumerate}
    \item \textbf{Step 1, State Preparation:} This step loads the uniform vector $\ket{\mathbf{b}}$ into the quantum circuit. To prepare this perfect superposition, we apply a single Hadamard gate to all $n_s$ system qubits. While this can be executed in $\mathcal{O}(1)$ depth, the total gate count is $\mathcal{O}(n_s) = \mathcal{O}(\log N)$. 
    \item \textbf{Step 2, 1-Bit Quantum Phase Estimation:} The QPE is implemented with the DSS method and a single phase rotation gate. The diagonal and off-diagonal components of $\mathbf{A}$ are dynamically loaded into the quantum circuit using the single phase rotation and DSS methods respectively. The DSS method iterates $k$ times and each iteration synthesizes an exact two-level unitary $G_k$ using three sub-routines:
    \begin{enumerate}
        \item \textbf{Basis Transformation:} A ladder of \texttt{CNOT} gates is applied to map the interacting states $\ket{i}$ and $\ket{j}$ such that they differ at only a single qubit index (the pivot), seen in Fig~\ref{fig:combined_circuit:c}. The number of two-qubit gates scales with the Hamming distance $d_H$ as $\mathcal{O}(d_H)$. Since $d_H \leq n_s$, this is upper-bounded by $\mathcal{O}(n_s)$.
        
        \item \textbf{Control Logic:} To trigger the multi-controlled rotation, $X$ gates are applied to any $n_s-1$ non-pivot qubits that are currently in the $\ket{0}$ state. This ensures that the multi-controlled rotation will target the subspace defined by $\{ \ket{i}, \ket{j} \}$ correctly. This step contributes $\mathcal{O}(n_s)$ single-qubit gates.
        
        \item \textbf{Rotation:} A multi-controlled $R_X(\theta)$ rotation is applied to the pivot qubit, controlled by the $n_s-1$ other qubits. Using the decomposition for multi-controlled rotations without ancilla \cite{Barenco_1995}, this contributes $\mathcal{O}(n_s)$ gates.
        
        \item \textbf{Uncomputation:} The $X$ gates and \texttt{CNOT} ladder are applied in reverse.
    \end{enumerate}
    Summing these components, a single DSS iteration scales linearly with the register size $\mathcal{O}(n_s)$. For $k$ iterations, the total complexity for Step 2 is $\mathcal{O}(k \cdot n_s)$.

    \item \textbf{Step 3, Controlled Inversion:} The controlled inversion consists of two $X$ gates and a single \texttt{CNOT} gate. This is constant $\mathcal{O}(1)$ complexity. 
    
    \item \textbf{Step 4, Uncomputation:} This is identical to Step 2, thus doubling the complexity contribution of the phase estimation. 
\end{enumerate}

The complexity is primarily determined by the $k$ repetitions required to encode every valid interaction term. Substituting $n_s = \log N$ and the sparsity factor $k \approx \sqrt{N}$, the total gate complexity is:
\begin{equation} 
C_{total} \approx k \cdot n_s \approx \mathcal{O}(\sqrt{N} \log N ) \; .
\label{eq:gate_complexity}
\end{equation}
This scaling represents a polynomial speedup in $N$ compared to classical exhaustive search methods\cite{RevModPhys.82.1419exaustive_search}, providing a computational advantage analogous to Grover's algorithm~\cite{Grover}.

\subsubsection*{Sampling Complexity}

Unlike the standard HHL algorithm, where the probability of measuring the solution state is suppressed by the condition number $\kappa$ as $\mathcal{O}(1/\kappa^2)$ \cite{Harrow_2009}, the performance of this 1-Bit Quantum Filter is governed by the geometric overlap of the initial state with the solution subspace. The algorithm succeeds when the ancilla is measured in the $\ket{1}_A$ state, which requires the time register to collapse to $\ket{0}_T$. The total success probability is the sum of the conditional probabilities for each eigenvector $\ket{u_j}$ weighted by its overlap with the input vector $\ket{\mathbf{b}}$:
\begin{equation}
    P_{\text{succ}} = P(T=0) = \sum_{j} P(0 | \lambda_j) \cdot |\gamma_j|^2 \; .
    \label{eq:succ}
\end{equation}
The Hamiltonian spectrum is partitioned into two orthogonal subspaces: the Signal subspace (valid tracks, $\lambda \neq \lambda_c$) and the Noise subspace (inactive segments, $\lambda = \lambda_c$). Substituting the filter response from Eq.~\eqref{eq:P1_succ}, the success probability becomes:
\begin{equation}
    P_{\text{succ}} \approx \frac{1}{2} \underbrace{\cos^2\left(\frac{\lambda_j \pi}{2 \lambda_c}\right)}_{\text{Signal Pass Rate}} \sum_{j \in \text{Signal}} |\gamma_j|^2 +   \frac{1}{2} \underbrace{\sin^2\left(\frac{\lambda_c \pi}{2 \lambda_c}\right)}_{\text{Noise Nulling}} \sum_{c \in \text{Noise}} |\gamma_c|^2  \; .
    \label{eq:p_succ}
\end{equation}
The filter deterministically eliminates the noise subspace (as the second term in Eq. \eqref{eq:p_succ} vanishes), while passing the signal component with an effective probability $c_{filter} = \cos^2\left(\frac{\lambda_j \pi}{2 \lambda_c}\right)$. The summation term $\sum_{j \in \text{Signal}} |\gamma_j|^2$ represents the overlap of the uniform superposition of $\ket{b}$ with the valid track configurations. For a system with $N$ total segments and $k$ valid interaction terms, this overlap is the ratio of the subspace
dimensions. Thus, the single-shot success probability scales linearly with the density of valid tracks, given $M = m(l - 1)$ valid track segments:
\begin{equation}
    P_{succ} \approx c_{filter} \cdot \sum_{j \in \text{Signal}} |\gamma_j|^2 \approx \frac{M \cdot c_{filter}}{N} \; .
    \label{eq:sampling_prob}
\end{equation}

To derive the full sampling complexity we account for the reconstruction of the entire event. Since the output is probabilistic, this process can be modeled by the Coupon Collector's Problem~\cite{Motwani_Raghavan_1995}, where we sample with replacement from $M$ distinct outcomes. To observe all distinct segments with high probability, the required number of successful measurements scales as $S_{\text{dist}} \approx M \log M$~\cite{Motwani_Raghavan_1995}. The total computational cost, defined as the total number of circuit executions $S_{\text{total}}$, is the required number of successes normalized by the single-shot probability $P_{\text{succ}}$:
\begin{equation} 
S_{\text{total}} = \frac{S_{\text{dist}}}{P_{\text{succ}}} \approx \frac{M \log M}{c_{\text{filter}}(M/N)} \approx N \log M \approx \mathcal{O}(N \log \sqrt{N}) \approx \mathcal{O}(N \log {N}) \; . 
\end{equation}
By expressing the track density in terms of the total system size $N$, we observe that the complexity scales as $\mathcal{O}(N \log {N})$. This represents an advantage over HHL, as we eliminate the polynomial dependence on the condition number $\kappa$\cite{Harrow_2009} and the precision overhead of the QPE. However, this does not overcome the fundamental challenge of quantum state tomography. To address this we would need to project onto more global event observables such as the particle's origin point (the "vertex") or employ some form of amplitude amplification on the $\ket{1}_A$ subspace.

\section*{Results}

\subsection*{Performance Evaluation}

To strictly evaluate the efficacy of the proposed 1-Bit Quantum Filter, we applied the algorithm to simulated event data generated with the simulation tool. This study centers on two critical performance metrics: (1) the ability to find the correct segments under realistic quantum noise conditions, with low fake rates, and (2) the empirical validation of the theoretically derived sampling and two-qubit gate complexities.

For these benchmarks, we employed a controlled version of the simulation tool. To study the scaling of the quantum algorithm we restricted the toy model to a single primary vertex with no detector inefficiencies or multiple scattering. We then varied the detector geometry between 3 and 5 layers to study the behavior and scaling of the $c_{filter}$ coefficient.  

To run our 1-Bit Quantum Filter, we utilized Quantinuum's TKET compilation stack~\cite{Sivarajah_2020_tket} and the Quantinuum Nexus platform~\cite{Quantinuum_Qnexus} to perform noiseless simulations, noise-model emulations, and hardware executions targeting the Quantinuum System Model H2 Quantum Processing Unit (QPU), which utilizes high-connectivity trapped-ion hardware. These results are compared against IBM’s Qiskit framework~\cite{javadiabhari2024quantumcomputingqiskit}, utilizing both the IBM Pittsburgh device and its associated noise model. This system represents the Heron R3 superconducting architecture~\cite{AbuGhanem_2025_ibm_heron}.

\begin{figure}[htb]
	\centering
    \includegraphics[width=17.0cm]{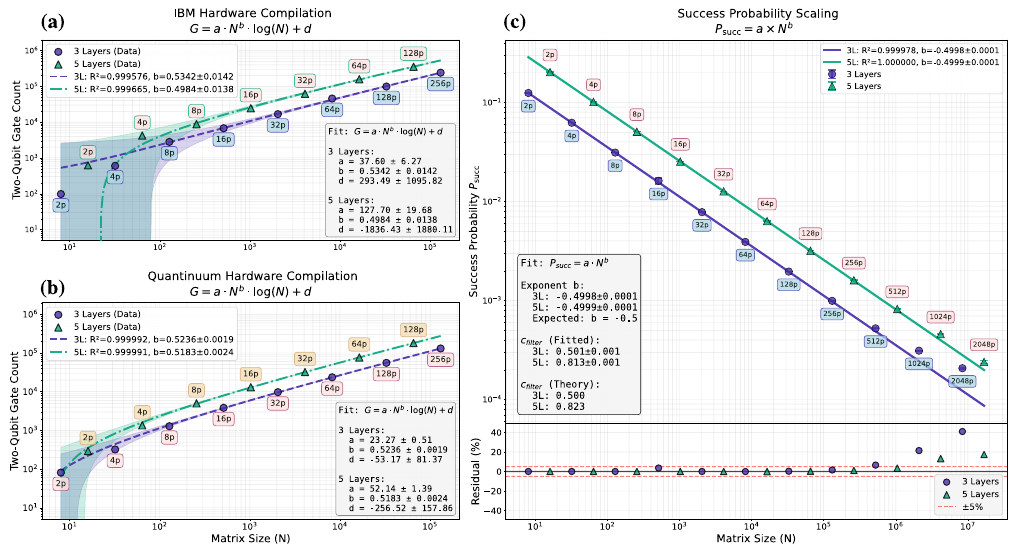}
    \caption{\textbf{Fig. 4: Asymptotic complexity scaling.} \textbf{(a)} Gate count complexity scaling for one-bit quantum filter implementations under IBM Pittsburgh hardware compilation. \textbf{(b)} Gate count complexity scaling under Quantinuum system model H2. Power-law fits with constrained logarithmic scaling ($c = 1.0$) yield scaling exponents of $b \approx 0.53$ for IBM and $b \approx 0.52$ for Quantinuum, consistent with the $\mathcal{O}(\sqrt{N} \log N)$ theoretical gate complexity ($R^2 > 0.999$). The three-layer tracking configurations are represented by blue dotted lines with circle markers, and the five-layer configurations are represented by green dotted lines with triangle markers. In all panels, the text labels annotated next to the data points (e.g., 4p, 128p) indicate the number of particles simulated to generate the corresponding matrix size.\textbf{(c)} Success probability $P_{\mathrm{succ}}$ as a function of matrix size $N$ for three-layer (blue solid line with circle markers) and five-layer (green solid line with triangle markers) tracking scenarios. The fitted exponents exactly match the theoretical $N^{-1/2}$ scaling up to $N = 10^6$, where an upward deviation is observed in the residuals. The error bars represent one standard deviation of the mean, calculated across a sample size of $n = 10$ independent statevector simulations, each utilizing $10^7$ measurement shots. The fitted scaling coefficient $c_{\mathrm{filter}}$ matches the theoretically predicted value for both the three-layer and five-layer configurations.}
    \label{fig:fits}
    \phantomsubcaption\label{fig:fits:a}
    \phantomsubcaption\label{fig:fits:b}
    \phantomsubcaption\label{fig:fits:c}
\end{figure}

\subsection*{Validation of Derived Complexities} 
\label{sec:validation_complexities}

To verify the theoretical two-qubit gate complexity of $\mathcal{O}(\sqrt{N} \log N)$ (Eq.~\ref{eq:gate_complexity}), we analyzed the post-compilation two-qubit gate counts for circuits targeting both IBM and Quantinuum architectures. Using the hardware compilation pipelines established above, we generated the necessary circuits using noiseless statevector simulations with Qiskit’s Aer simulator\cite{javadiabhari2024quantumcomputingqiskit}, accelerated via an Nvidia L40S GPU. For the compilation step, we specifically applied Qiskit’s optimization level 3 for the IBM Pittsburgh backend, and ‘full peephole’ optimization\cite{Sivarajah_2020_tket} within the TKET framework. The gate count analysis is restricted to the range $N \in [2^3, 2^{17}]$ , as the classical optimization of circuits exceeding $10^6$ two-qubit gates becomes computationally intractable beyond this point.

The compiled gate counts are fitted to the theoretical ansatz $C_{total} = a N^b \log(N) + c$. The logarithmic term is fixed to reflect the dependence on register size ($n_s = \log N$), allowing us to isolate the polynomial scaling exponent $b$. This constraint minimizes the degrees of freedom relative to the dataset size preventing overfitting, in the asymptotic regime.

Figures \ref{fig:fits:a} and \ref{fig:fits:b} display the results with $1 \sigma$ confidence bands. We observe excellent agreement between the fitted coefficient $b$ and the theoretical prediction of $0.5$ in the asymptotic regime ($N > 2^{10}$), where the confidence bands narrow significantly. The divergence observed at smaller system sizes is attributed to the constant-depth overhead not captured by the asymptotic model. 

Comparing the two compilation pipelines reveals a distinct scaling behavior driven by hardware topology. The TKET results (Fig.~\ref{fig:fits:b}), which target the all-to-all connectivity of trapped-ion systems, successfully minimize the overall gate count as the hardware supports direct interaction between arbitrary qubit pairs removing the necessity for SWAP gates. This topological advantage is explicitly reflected in the fitted parameters, yielding tight confidence intervals, lower scaling pre-factors (e.g., $a = 23.27 \pm 0.51$ for 3 layers), and minimal constant overhead ($d$). In contrast, Qiskit results (Fig.~\ref{fig:fits:a}) exhibit larger overhead and higher variance. This discrepancy arises from the limited connectivity of the IBM superconducting architecture, which requires the insertion of SWAP chains during the routing process to implement non-local Hamiltonian terms. This complex routing inflates both the scaling coefficient $a$ rising to $a = 37.60 \pm 6.27$ for 3 layers and $a = 127.78 \pm 19.68$ for 5 layers.

Having established the theoretical total sampling complexity in Eq.~\eqref{eq:sampling_prob}, we see that the validation of our model is entirely dependent on the behavior of the success probability $P_{succ}$ with system size $N$, as $S_{dist}$ follows the standard  coupon collector scaling. Rewriting Eq.~\eqref{eq:p_succ} in terms of the system size $N$, the theoretical success probability is expected to scale as $P_{succ} \approx (c_{filter}\sqrt{l-1}) / \sqrt{N}$. As such, we fit the simulation data to a power-law model of the form $P_{succ} = a \cdot N^b$. Validation of the theoretical derivation requires that the fitted exponent $b$ converges to $-0.5$ and the coefficient $a$ aligns with the derived factor $c_{filter} \sqrt{l-1}$.

The dataset for this scaling analysis was generated using statevector simulations. Since determining $P_{succ}$ does not require the computationally expensive hardware specific compilation step used in the gate analysis, we extended the range from the minimal configuration of 3 layers and 2 tracks ($N=2^3$) up to a denser environment of 5 layers and 1024 tracks ($N=2^{22}$). Experimentally, $P_{succ}$ is quantified as the probability of the ancilla flag qubit projecting onto the $\ket{1}$ subspace. Figure \ref{fig:fits:c} presents the power-law fits for 3-layer and 5-layer detector configurations. The fitted coefficients $a$ and exponents $b$ exhibit excellent agreement with theoretical predictions for both configurations ($R^2 > 0.999$). Specifically, the fitted exponents match the expected $N^{-0.5}$ scaling ($b = -0.4998 \pm 0.0001$ for 3 layers, and $b = -0.4999 \pm 0.0001$ for 5 layers). Furthermore, the empirical pre-factors yield $c_{filter}$ values of $0.501 \pm 0.001$ and $0.813 \pm 0.001$ for 3 and 5 layers respectively, tightly aligning with our theoretical $c_{filter}$ derivations of $0.500$ and $0.823$. This confirms the validity of the derived scaling laws. 

A deviation is observed for large system sizes ($N > 2^{18}$), where the measured $P_{succ}$ exceeds the theoretical prediction. This enhancement is not a computational artifact but a consequence of the combinatorial density. In regions of high track occupancy, the density of hits increases the probability of fake segments satisfying the angular constraints of the Hamiltonian Eq.~\eqref{eq:angular_term}. Because the current implementation relies on a simple binary threshold defined in Eq.~\eqref{eq:angular_term}, random geometric alignments can satisfy the angular constraints. These false positives satisfy the Hamiltonian constraints and contribute to the final solution state, increasing the $P_{succ}$ even though they do not correspond to any real tracks.  This was also observed using classical matrix inversion in the simulation tool, and could be resolved by adding additional Denby-Peterson terms to the Hamiltonian.

\subsection*{Benchmarking Realistic Event Topologies}

\begin{figure}[htb]
	\centering
    \includegraphics[width=17.0cm]{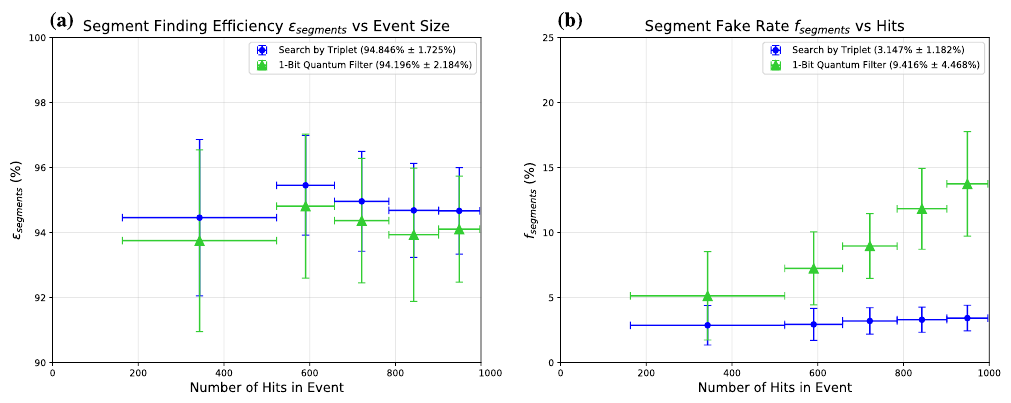}
    \caption{\textbf{Fig. 5: Monte Carlo Performance comparisons.} The performance of our 1-Bit Quantum Filter simulation (green triangles) on realistic LHCb simulated (MC) events, is benchmarked against the classical state-of-the-art Search by Triplet\cite{search_by_triplet} algorithm (blue circles) across varying event sizes. The horizontal error bars represent equal density bins and the vertical error bars represent a 1 standard deviation of the mean in that bin for which 79 events were used per bin.\textbf{(a)} Segment Finding Efficiency ($\epsilon_{segment}$ Eq.~\ref{eq:segment_eff}), representing the fraction of true track segments successfully recovered as a function of the total number of detector hits. \textbf{(b)} Segment fake rate ($f_{segment}$ Eq.~\ref{eq:segment_eff}), representing the fraction of incorrectly identified segments, as a function of the total number of detector hits.}
\label{fig:mc_quantum_simulation}
\phantomsubcaption\label{fig:mc_quantum_simulation:a}  
\phantomsubcaption\label{fig:mc_quantum_simulation:b}

\end{figure}

While the algorithm demonstrates theoretical scaling advantages, evaluating its robustness against realistic detector conditions is critical to assessing its practical relevance. To this end, we utilized quantum statevector simulations to process realistic LHCb Monte Carlo (MC) events. Specifically, we utilized the same $B_s \rightarrow \phi \phi$ selection that has been previously used to obtain tracking efficiencies using classical inversion \cite{Nicotra_2023}. The $B_s \rightarrow \phi \phi$ events were generated from $pp$ collisions at a center-of-mass energy $\sqrt{s}=13.6 \; \text{TeV}$, corresponding to LHC run 3 and 4. The event generation and detector response were simulated using the standard combination of Pythia\cite{Torbjorn_Sjostrand_2006_pythia} with a specific LHCb configuration\cite{Belyaev_2011_lhcb_config}, EvtGen\cite{LANGE2001152_EvtGen} for hadronic decays, and Geant4\cite{AGOSTINELLI2003250_Geant4} to simulate the material interactions\cite{Clemencic_2011_interation_material}. We select events with up to 1000 hits in the right half of the VELO. The event is then processed using a quantum simulation of the 1-Bit Quantum Filter algorithm, which allows us to probe the algorithm's response in tracking environments with realistic detector inefficiencies. 

We benchmarked the performance of our 1-Bit Quantum Filter simulation directly against the algorithm that is currently used in LHCb reconstruction, the Search by Triplet (SBT) algorithm \cite{search_by_triplet}, utilizing its open-source Python implementation~\cite{campora_velopix_repo}. To ensure a direct and fair comparison with the distribution sampled from the quantum simulation, performance was evaluated strictly at the segment level, using the segment finding efficiency $\epsilon_{segments}$ and the fake rate $f_{segments}$ as defined in Eq.~\ref{eq:segment_eff}.

As seen in Fig.~\ref{fig:mc_quantum_simulation:a}, the 1-Bit Quantum Filter remains effective in realistic events with up to 1,000 hits. The quantum algorithm achieved a mean $\langle \epsilon_{segments} \rangle = 94.2\% \pm 2.2\%$ across the tested event sizes, which is close to the SBT's efficiency of $\langle \epsilon_{segments} \rangle = 94.8\% \pm 1.7\%$. However, examining the segment fake rate in Fig.~\ref{fig:mc_quantum_simulation:b} reveals a different trend. The SBT algorithm's $f_{segments}$ exhibits a minimal increase with event size, ranging from $2.9\%-3.4\%$, whereas the quantum algorithm yields a range of $5.1\%-13.7\%$.

Fundamentally, these false positive segments originate from the construction of our Hamiltonian, which currently relies solely on an angular step function. As discussed previously, suppressing these with Denby-Peterson penalty terms~\cite{DP1, DP2} is possible, but remains a subject for further study due to the necessary trade-off with increased quantum circuit depth.

\subsection*{Solution Fidelity on noisy hardware}

\begin{figure}[htb]
	\centering
    \includegraphics[width=17.0cm]{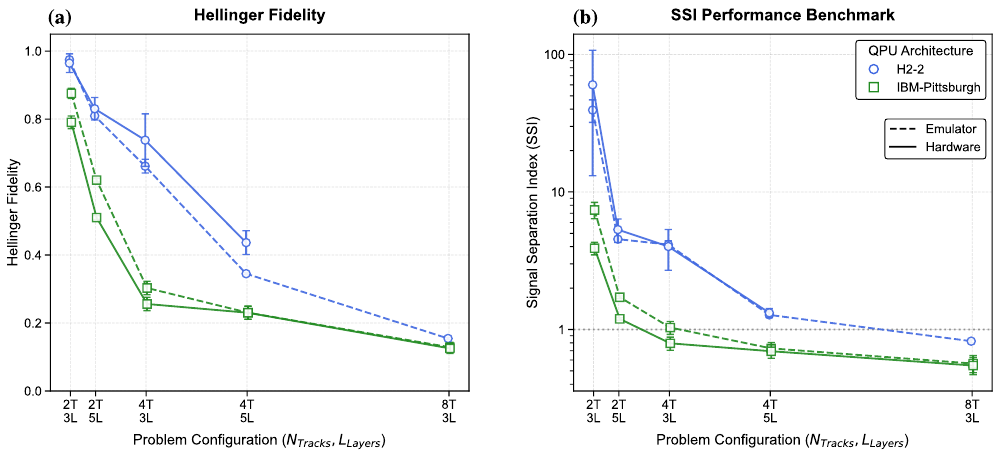}
    \caption{\textbf{Fig. 6: Hardware performance comparison.} Both panels present metrics evaluating tracking results for problem sizes up to 8 tracks across 3 layers, utilizing both quantum hardware and noise emulators. All results were obtained across the 5 tracking scenarios, except for the physical Quantinuum H2 hardware executions, which are limited to the first 4 scenarios due to deep circuits and constrained quantum resources. For IBM Pittsburgh all points were generated across 5 runs with 10,000 shots each. For system model H2 the points were generated across 5 runs with the first 3 using 500 shots and the 4th with 800 shots. Error bars denote $1\sigma$ confidence intervals calculated directly across independent algorithm executions. \textbf{(a)} Hellinger fidelity ($F_H$) relative to the noiseless baseline indicating the overlap of the probability distribution. \textbf{(b)} Signal Separation Index (SSI) quantifying the contrast between valid tracks and the noise floor. The dashed line at $\text{SSI}=1$ represents the distinguishability limit where signal amplitudes drop below the background noise.}
\label{fig:fidelity}
\phantomsubcaption\label{fig:fidelity:a}
\phantomsubcaption\label{fig:fidelity:b}
\end{figure}

The noise resilience of the proposed 1-bit quantum filter is assessed by executing the circuits on both physical quantum hardware and noise-model emulators for the IBM Pittsburgh (Heron R3) and Quantinuum H2-2 processor. The noise models are derived from regular device calibration data, accurately reflecting the coherence times, gate fidelities, and readout error rates of the physical hardware at the time of execution.We employ the same compilation pipeline established for the validation of our theoretical complexities. For the superconducting executions, the Qiskit compilation step was modified to target the specific IBM hardware backend, while for the trapped-ion executions, the optimized TKET circuits were compiled and executed via the Quantinuum Nexus platform. For the IBM hardware executions, baseline error mitigation was applied using Twirled Readout Error eXtinction (TREX)\cite{PhysRevA.105.032620TREX} via Qiskit's Resilience Level 1. The analysis is restricted to $N \in [2^3, 2^7]$ to ensure the accumulated gate errors remain within a regime where valid tracks can be resolved from the background. For statistical validation, executions on the physical IBM Pittsburgh processor, as well as all noise-model emulators, were performed using 10,000 measurement shots across 5 independent repetitions. However, physical executions on the Quantinuum H2-2 hardware were constrained by quantum resource availability, necessitating a reduced sampling scheme of 3 independent repetitions. Within these H2-2 hardware runs, the first three tracking configurations were evaluated using 500 shots, while the fourth configuration utilized 800 shots. Despite this reduction in sampling volume, the allocated shot counts proved statistically sufficient to reconstruct the final probability distributions and extract performance metrics.

To quantify the impact of noise on track reconstruction performance we construct the following two metrics: 

\begin{enumerate}
    \item \textbf{Hellinger Fidelity ($F_H$)}: Measures the geometric overlap between the noisy experimental distribution $P_{exp}$ and the ideal noiseless baseline $P_{ideal}$:
    \begin{equation}
    F_{H}(P_{exp}, P_{ideal}) = \left( \sum_{i} \sqrt{P_{exp}(x_i) \cdot P_{ideal}(x_i)} \right)^2
    \end{equation}
    \item \textbf{Signal Separation Index (SSI)}: This metric quantifies the distinguishability of valid tracks from the background. We define it as the ratio of the total probability mass of the $M$ valid track segments (${S}$) to that of the $M$ highest-amplitude noise segments (${N}^*$):
    \begin{equation}
    \text{SSI} = \frac{\sum_{i \in {S}} P_i}{\sum_{j \in {N}^*} P_j}
    \end{equation}
\end{enumerate}

The Hellinger Fidelity ($F_H$) results, presented in Fig.~\ref{fig:fidelity:a}, demonstrate a distinct performance hierarchy among the evaluated architectures, a trend that is consistent across both the physical hardware executions and the noise emulators. The Quantinuum H2 device consistently achieves the highest fidelity across all problem configurations when compared to the IBM Pittsburgh device.  Furthermore, distinct deviation trends emerge between the physical executions and their emulators. For the IBM architecture, the emulator consistently predicts a slightly higher fidelity than the physical hardware achieves. Conversely, the Quantinuum H2-2 device demonstrates tight agreement between hardware and emulator for small configurations, while the physical hardware actually outperforms the emulator at larger problem sizes. 
The observed difference in performance between hardware devices may be caused by the topological differences between the architectures. The all-to-all connectivity of the trapped-ion H2 device allows the compiler (TKET) to synthesize the unitary with approximately half the circuit depth required for the fixed-topology superconducting chips. Since error accumulation is exponential with depth, this topological advantage and higher qubit fidelity\cite{Moses_2023_race_track} enhances the solution quality.

The Signal Separation Index (SSI), shown in Fig.~\ref{fig:fidelity:b}, corroborates these findings by mapping the distinguishability of valid tracks from fake segment combinations. The dashed horizontal line at $\text{SSI}=1$ defines the critical threshold below which the quantum filter fails to separate signal from noise. When comparing the physical executions to their respective emulators, the SSI trends mirror those observed in the Hellinger fidelity. The IBM Pittsburgh hardware performance is below its emulator predictions whereas the Quantinuum H2 hardware and emulator demonstrate agreement, although with larger statistical uncertainties in the hardware data due to a limited number of measurement shots. Consistent with the fidelity data, both the hardware and emulator results for the H2 platform maintain a resolvable signal ($\text{SSI} > 1$) up to the denser 4-track, 5-layer configuration. In contrast, executions on hardware and emulators using the superconducting architecture approach the noise floor more rapidly. These results indicate that for current NISQ hardware, the effective connectivity and its direct reduction of compilation overhead may be a dominant factor in extending the reach of quantum track reconstruction to larger detector geometries.

\section*{Discussion}

In this paper we presented a 1-Bit Quantum Filter, a domain-specific adaptation of the HHL algorithm tailored for the combinatorial challenge of particle track reconstruction. Our results demonstrate that by replacing standard quantum linear algebra routines in favour of problem-informed filtering, we can achieve a polynomial gate complexity $\mathcal{O}(\sqrt{N} \log N)$.

A key finding of this study is the computational advantage gained by relaxing the requirements of the HHL algorithm, originally designed to perform precise matrix inversion $A^{-1}b$. This precision preserves the relative amplitudes of the solution vector for any well conditioned matrix, however, for the particle tracking task this is excessive. The physics problem is inherently binary: we aim to distinguish the "signal" subspace (valid track segments) from the "noise" subspace (combinatorial background segments), rather than computing a continuous amplitude distribution. This 1-Bit Quantum Filter exploits this relaxation by replacing the resource intensive multi-qubit QPE with a single-ancilla projection. The modification effectively replaces the expensive arithmetic inversion of the eigenvalues ($1/\lambda$) with a simpler spectral step function, significantly reducing the circuit depth and the number of required ancilla qubits. The shift from calculating the exact amplitudes of the solution spectrum to filtering only track segments is what renders the algorithm feasible for the problem sizes required by the HL-LHC, where the overhead of full QPE would be prohibitive. This scalability is further supported by previous studies demonstrating that the Hamiltonian matrix becomes increasingly sparse with increasing event sizes \cite{Nicotra_2023}. To contextualize this scalability, HL-LHC luminosities during LHCb Run-5 are expected to yield 1,000 to 2,000 charged particles per event. Due to the combinatorial density of hits, this multiplicity pushes the matrix size to $N \approx 10^6 - 10^7$ candidate segments.  In contrast to Search-by-Triplet\cite{search_by_triplet}, which is a local tracking algorithm, our Hamiltonian formulation operates as a global pattern recognition method. Integrating this global filtering into future heterogeneous, hybrid tracking pipelines could resolve complex combinatorial problems which will be seen at the HL-LHC.

While the gate complexity of our approach scales favorably, the total runtime remains constrained by the sampling complexity, which scales linearly with the system size $\mathcal{O}(N)$. To reconstruct the full event, the quantum state must be repeatedly prepared and measured until all $M$ valid track segments are found. For large events on the order of the HL-LHC the linear readout cost is expensive. Future research must therefore pivot from full state tomography to the direct estimation of global properties. Instead of reading out the full distribution, one could construct quantum observables that can extract higher level physics features that are easy to implement on hardware. One such feature is the ${z}$-coordinate location of the Primary Vertices (PVs). In high-energy physics, the PV represents the primary point of particle collision on the beamline and at the HL-LHC ~100 PVs may be present in an event. Their extraction at a low level reconstruction stage would greatly benefit complete event reconstruction and selection. By extracting this compressed information directly from the quantum state, it may be possible to bypass the $\mathcal{O}(N)$ track readout bottleneck entirely and achieve sub-linear readout costs.

The results obtained by running our 1-bit HHL algorithm in Quantinuum and IBM hardware show promising outcomes for small systems of about four particle tracks across three detector layers. Larger systems scale badly with size as separation of good and back track segments disappears due to buildup of gate errors and the noise floor of current quantum devices. The results in Fig.~\ref{fig:fidelity} on fidelity and Signal Seperation Index show a more favorable result for Quantinuum, which indicates that the all-to-all connectivity is better suited to our use case than a fixed topology requiring multiple SWAP operations. While our conclusions highlight current hardware constraints, future fault-tolerant quantum computers will use Quantum Error Correction (QEC) which may mitigate physical connectivity limits. How these codes will be implemented across architectures remains an open question. Ultimately, the algorithm's scalability will depend on balancing QEC overhead against the challenges of scaling physical qubits.

The benchmarks presented here evaluate our 1-Bit Quantum Filter on both physical quantum hardware and corresponding noise models. While the noise models capture general machine errors, our dual approach successfully accounts for the nuanced differences between simulated and physical noise across different hardware platforms. While current NISQ-era hardware restricts the solvable problem size, these physical executions experimentally validate our proof-of-principle. Future work will focus on benchmarking the algorithm within logical error-correction layers, optimizing both the Hamiltonian formulation and this 1-Bit Quantum Filter, the direct extraction of global event observables such as Primary Vertices, and the efficient calculation and construction of the Hamiltonian matrix.

\section*{Conclusion}

Our 1-Bit Quantum Filter represents a development towards a potential quantum-classical hybrid particle reconstruction algorithm, focusing towards specialized, problem-specific spectral filtering.
We have validated the theoretical two-qubit gate complexity $\mathcal{O}(\sqrt{N} \log N)$ and simulated our 1-Bit Quantum Filter on realistic LHCb Monte Carlo events, a task previously inaccessible by quantum tracking algorithms. We demonstrate that for small systems the algorithm is robust against noise profiles of current generation hardware, particularly on architectures with high connectivity. As quantum hardware matures from noisy physical qubits to logically coherent systems, this 1-Bit Quantum Filter deserves further study to reconstruct events at the scale of the future HL-LHC. More broadly, this 1-Bit Quantum Filter technique can be adapted to similar optimization problems such as the ones arising in the context of graph optimization, associative memories and pattern denoising, reducing the practical requirements down to sizes that are accessible by quantum hardware of today.

\section*{Data availability}

The source data used in this study for Fig.~\ref{fig:Event}, Fig.~\ref{fig:SimulationOverview}, Fig.~\ref{fig:fits}, Fig.~\ref{fig:mc_quantum_simulation}, Fig.~\ref{fig:fidelity} can be found in the data repository available at https://github.com/Xenofon-Chiotopoulos/OneBQF.

\section*{Code availability}

The code used in this work is open source and available at https://github.com/Xenofon-Chiotopoulos/OneBQF

\bibliography{sample}

\section*{Acknowledgements} 
We are thankful to the CERN Quantum Technology Initiative (QTI) for providing access to IBM Quantum computers and Quantinuum for providing access to their quantum platforms, resulting in the studies reported in this paper. We also thank the Coherence and Quantum Technology group of Eindhoven University for initial tests on their simulation platform. Finally, we are grateful for the support from the LHCb Data Processing and Analysis (DPA) project, and acknowledge the LHCb simulation and computing projects for the production of simulated samples.

\section*{Funding Statement}

X. Chiotopoulos discloses support for the research of this work from the NWO project {\it Fast Sensors and Algorithms for Space-time Tracking and Event Reconstruction (FASTER)} [grant number OCENW.XL21.XL21.076], D. Nicotra discloses support for part of the research from SURF innovation fund: {\it From Quark to Quantum} [SOIL.QQLHCb.01], and G. Scriven discloses support for the research of his work by the Special Research Fund of Hasselt University [grant number BOF24DOCUM01]. K. Driessens, M. Merk, J. Sch\"{u}tz, J. de Vries and M. Winands declare no relevant funding.

\section*{Author contribution}

\subsection*{Authors and Affiliations}
\textbf{Nikhef National Institute for Subatomic Physics, Science Park 105, 1098 XG Amsterdam, The Netherlands}

\vspace{0.1cm}
Xenofon Chiotopoulos, Marcel Merk
\vspace{0.1cm}

\textbf{Maastricht University, Faculty of Science and Engineering, Gravitational Waves and Fundamental Physics department, Duboisdomain 30, 6229 GT Maastricht, The Netherlands}

\vspace{0.1cm}
Xenofon Chiotopoulos, George Scriven, Davide Nicotra, Jacco de Vries, Marcel Merk
\vspace{0.1cm}

\textbf{{Hasselt University, Faculty of Sciences and Data Science Institute, Agoralaan gebouw D, 3590 Diepenbeek, Belgium}}

\vspace{0.1cm}
George Scriven, Jochen Sch\"{u}tz
\vspace{0.1cm}

\textbf{Maastricht University, Faculty of Science and Engineering, Department of Advanced Computing Sciences,  Paul-Henri Spaaklaan 1, 6229 EN Maastricht, The Netherlands}

\vspace{0.1cm}
Xenofon Chiotopoulos, Kurt Driessens, Mark H.M. Winands

\subsection*{Contributions}

X.C. derived and implemented the 1-Bit Quantum Filter and the Direct Structural Synthesis (DSS) method, performed the theoretical and empirical data fitting complexity analysis with cross checks from D.N., performed the hardware and emulator executions on both platforms, and carried out the tracking benchmarking on LHCb Monte Carlo events. X.C. M.M., G.S., D.N., K.D., M.H.M.W. and J.d.V. contributed to the interpretation of these benchmarks. G.S. developed the simulation framework based on original work by D.N., updated the HHL implementation to the latest Qiskit version, and together with M.M. developed the toy simulation model. J.d.V., K.D., M.M. and D.N. involved in the discussions of the Hamiltonian formulation. J.S. contributed to the mathematical consistency of the work. All authors participated in the theoretical and algorithm discussions, contributed to the interpretation of the results and edited the manuscript.

\subsection*{Corresponding author}
Correspondence to Xenofon Chiotopoulos (xenofon.chiotopoulos@maastrichtuniversity.nl)

\section*{Competing interests}
The authors declare no competing interests.

\end{document}